\documentclass{pasj01}
\Received{$\langle$reception date$\rangle$}
\Accepted{$\langle$acception date$\rangle$}
\Published{$\langle$publication date$\rangle$}

\newcommand{\rxj}{\hbox{RX J1713.7$-$3946}}
\newcommand{\rxjG}{\hbox{G347.3$-$0.5}}

\newcommand{\mass}{\mu_{\rm H}}

\usepackage{cases}

\begin{document}

\title{Expansion measurement of Supernova Remnant RX J1713.7$-$3946}
\author{Naomi \textsc{Tsuji} and Yasunobu \textsc{Uchiyama} }
\affil{Department of Physics, Rikkyo University, 
 3-34-1 Nishi Ikebukuro, Toshima-ku, Tokyo 171-8501, Japan}
\email{n.tsuji@rikkyo.ac.jp}
\email{y.uchiyama@rikkyo.ac.jp}


\maketitle

\begin{abstract}
Supernova remnant (SNR) \rxj\ is well known for its bright TeV gamma-ray emission with shell-like morphology. Strong synchrotron X-ray emission dominates the total X-ray flux in SNR \rxj\ and 
the X-ray morphology is broadly similar to the TeV gamma-ray appearance. 
The synchrotron X-ray and TeV gamma-ray brightness allows us to perform detailed analysis 
of the acceleration of TeV-scale particles in this SNR. 
To constrain the hydrodynamical evolution of \rxj, 
we have performed six times observations of the northwestern (NW) shell with the {\it Chandra} X-ray Observatory from 2005 to 2011, and measured the proper motion by using these data and the first epoch observation taken in 2000. 
The blast-wave shock speed at the NW shell is measured to be $(3900\pm 300) (d/{\rm kpc})\ {\rm km}\ {\rm s}^{-1}$ with an estimated distance of $d = 1$ kpc, and the proper motions of other structures within the NW shell are significantly less than that. 
Assuming that the measured blast-wave shock speed is the representative of the remnant's outer shock wave as a whole, 
we have confronted our measurements as well as  a recent detection of thermal X-ray lines,  with 
 the analytic solution of the hydrodynamical properties of SNRs. 
Our hydrodynamical analysis indicates that the age of the remnant is 1580--2100 years,  
supporting the association with SN393. 
A model with SN kinetic energy of $E = 10^{51}\ \rm erg$, the ejecta mass of 
$M_{\rm ej} = 3\, M_{\Sol}$, and the ambient density 
at the current blast wave location of $n_2 = 0.015\ {\rm cm}^{-3}$, provides reasonable 
explanation for our measurements and previous findings at the X-ray and gamma-ray 
wavelengths. 
We find that the transition to the Sedov-Taylor (ST) phase is incomplete for any reasonable set of parameters, implying that the current maximum energy of accelerated protons in \rxj\ would not correspond to the maximum attainable energy for this remnant.
\end{abstract}



\begin{figure*}[ht]
 \begin{center}
 \includegraphics[height=7cm]{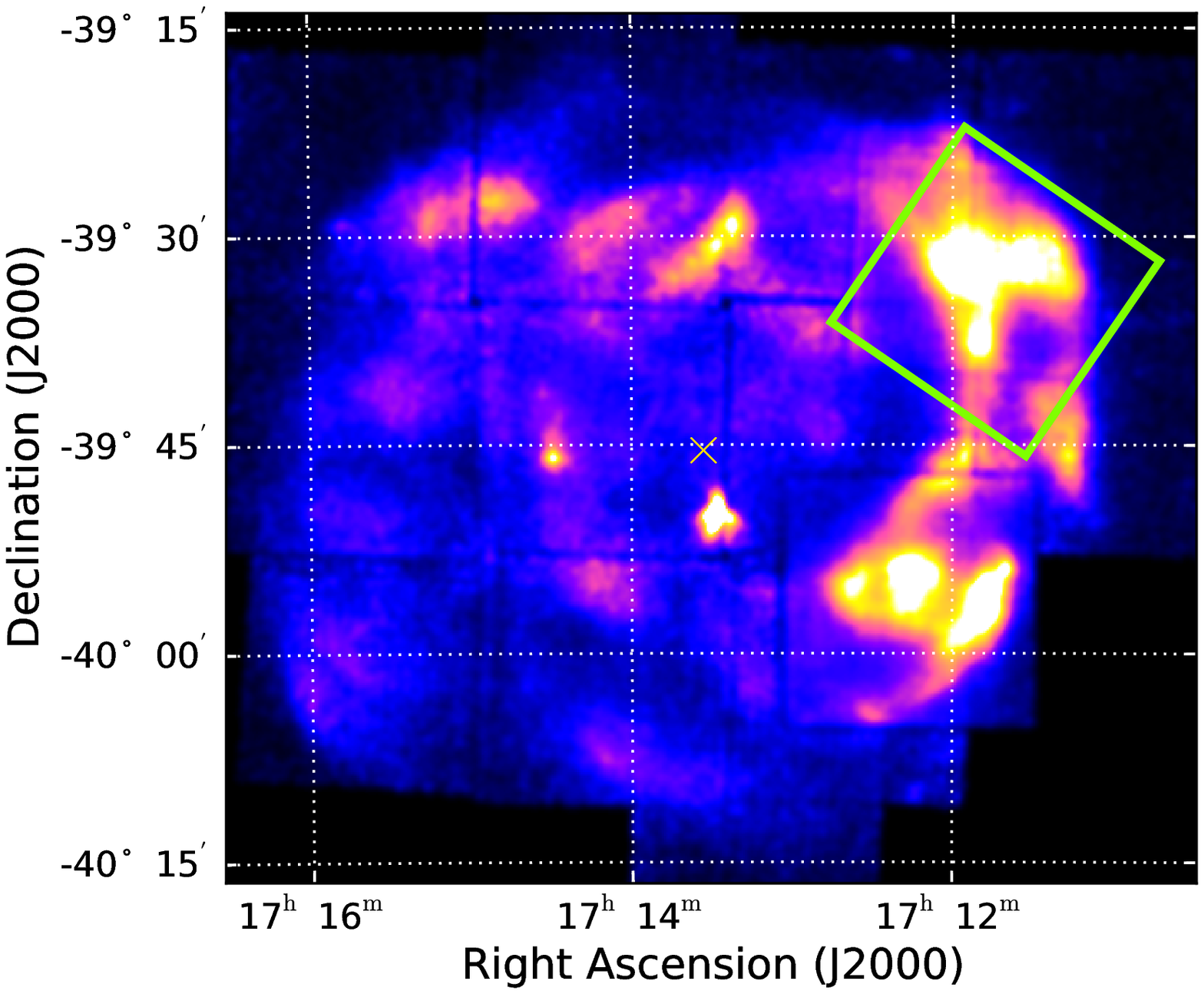} 
 \includegraphics[height=7cm]{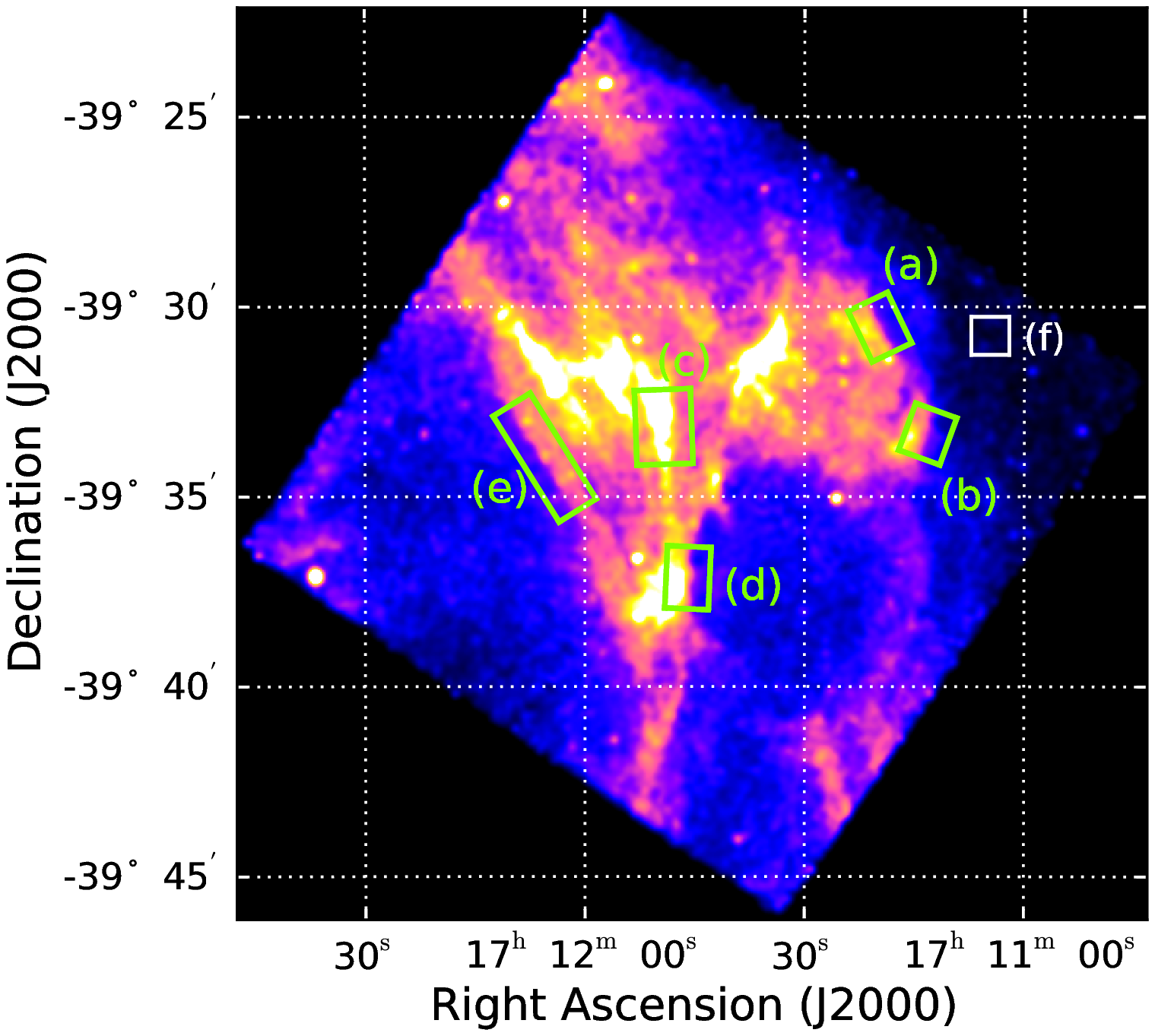} \\
  \end{center}
 \caption{{\it Left panel}: The flux image covering the entire region of SNR \rxj\ in 0.5$-$10.0 keV observed by the {\it Suzaku} XIS. 
 The green square region shows the FoV of the {\it Chandra} ACIS-I observations in 2011. 
 The yellow cross indicates the nominal center of the remnant: \timeform{17h13m33.6s}, \timeform{-39D45'36''} (J2000). 
 {\it Right panel}: The 0.5$-$5.0 keV flux image of the northwestern region of SNR \rxj\ taken by the {\it Chandra} ACIS-I in 2011. 
 Boxes named  (a) to  (e) are the regions where we measure the proper motion. Box (f) is the representative of the faint regions outside \rxj.}
 \label{fig:box}
\end{figure*}

\section{Introduction}

The shell-type supernova remnant (SNR) \rxj\ (also known as \rxjG) was discovered as an extended X-ray source with the angular size of about 60\arcmin$\times$50\arcmin, centerd at \timeform{17h13m42s}, \timeform{-39D46'27''} (J2000) by {\it ROSAT} X-ray all-sky survey \citep{Pfeffermann1996}. The distance from Earth to this SNR is estimated to be about 1 kpc from the association with molecular clouds traced by 
CO lines \citep{Fukui2003, Fukui2012, Sano2013}. 
\citet{Wang1997} proposed that \rxj\ is a remnant of the Chinese ``guest" star in 393 AD (hereafter SN393) that appeared within the curled tail of Scorpius according to the historical descriptions. 
However, \citet{Fesen2012} argued that  SN393 should have been very bright with the peak apparent visual brightness of 
$m_{\rm v} = -3.8$ to $-5.2$ and visible for over a year, which would be incompatible with the  
the Chinese literature who did not mention the apparent brightness and stated that the guest star disappeared after eight months. 

Putting aside the possible association with SN393, 
 \rxj\ is widely regarded  as a relatively young supernova remnant due to its strong nonthermal X-ray emission. 
 The nonthermal X-ray emission from \rxj\ is thought to be synchrotron radiation produced by high energy electrons 
 (\cite{Koyama1997}; \cite{Slane1999}; \cite{Tanaka2008}) that are 
 accelerated presumably via diffusive shock acceleration (DSA: \cite{Axford1977}; \cite{Blandford1978}; \cite{Krymskii1977}; \cite{Bell1978}). 
  
SNR \rxj\ is one of the most important objects for the study of the shock acceleration 
machanism thanks not only to its strong nonthermal X-ray emission but also to the detection of high energy and very-high energy gamma rays (\cite{Abdo2011}; \cite{Aharonian2004}, 2006, 2007). 
The image of the TeV gamma-ray emission is broadly similar to the synchrotron X-ray image. 
The main radiation process responsible for the observed gamma-ray emission
has been matter of active debate (e.g., \cite{Uchiyama2003}; \cite{Aharonian2006}; \cite{Berezhko2006};   \cite{Tanaka2008}; \cite{Ellison2010}; \cite{Zirakashvili2010}; \cite{Abdo2011}; \cite{Fukui2012}; \cite{Inoue2012}, \cite{Sano2013}); the gamma rays can be produced by Inverse Compton (IC) scattering of high energy electrons  (``leptonic model") or by neutral pion decay resulting from the collisions of high energy protons and ambient protons (``hadronic model").

In the framework of simple one-zone modeling where high-energy electrons and 
protons occupy the same volume with constant density and magnetic field strength, 
the hard spectrum in the GeV gamma-ray band measured with the {\it Fermi} Large Area Telescope (LAT) would be better described by the leptonic model \citep{Abdo2011}.
The magnetic field strength is estimated to be $\sim10\ \mu{\rm G}$ from 
 the flux ratio between synchrotron X-ray and gamma-ray radiation 
within this framework. 
The magnetic field, however, is probably enhanced (amplified) up to $\sim0.1\mbox{--}1\ {\rm mG}$ at some local scales, as manifested by 
the filamentary structures of synchrotron X-ray emission as well as the year-scale X-ray variability \citep{Uchiyama2007}.  The strong local magnetic field would be related to 
the interaction with molecular clouds \citep{Fukui2003, Inoue2012, Sano2013}.
On the other hand,  if we introduce highly inhomogeneous medium such as the presence of dense clumps in a low density cavity \citep{Inoue2012}, 
the gamma-ray emission would also be explained by the hadronic model \citep{Zirakashvili2010}. 
\citet{GA2014} demonstrated that the hard GeV spectrum can be fit by $\pi^0$-decay gamma rays that takes account of energy-dependent penetration of high-energy protons into dense clumps.

The measurements of synchrotron radiation and gamma-ray radiation do not provide sufficient information of 
the hydrodynamical properties of the remnant. Instead, one needs to measure thermal radiation in order to constrain the physical parameters. 
After the long term to struggle with finding thermal components from \rxj, \citet{Katsuda2015} has recently discovered thermal X-ray emission around the central region of \rxj. Based on the much higher abundance ratio of measured Ne, Mg and Si to Fe than the solar value, the progenitor mass is inferred to be less than $\sim 20\ M_{\Sol}$. 
Still, most of the physical parameters of SNR \rxj\ are ambiguous. Particularly it is quite meaningful to measure the age of SNR \rxj\ and to clarify which phase it is in, i.e. ejecta-dominated (ED) stage (free expansion stage) or Sedov-Taylor stage (ST stage).
For example, the maximum energy of the accelerated protons at SNR shock waves, $E_{\rm max}$, is expected to reach its maximum 
 around  the transition time from the ED stage to the ST stage because DSA theory generally predicts $E_{\rm max}$ decreases with time 
 during the ST stage: e.g., $E_{\rm max} \propto t^{-\mu-1/5}$ 
 for $B \propto t^{-\mu}$ \citep{Caprioli2010}. 


In this paper, we investigate the hydrodynamical properties based on  the X-ray expansion measurement of \rxj.  In \S\ref{sec:Observation}, we present the observation data sets. Section \ref{sec:sec3} describes the method of measuring the proper motions of X-ray emitting shells. Section \ref{sec:Models} presents hydrodynamical model to describe the evolution of the SNR. In \S\ref{sec:discussion}, we discuss which evolutional model has good agreement with the result of the expansion measurement obtained in \S\ref{sec:sec3}, using the analytic solutions. 
Finally,  conclusions are given in \S\ref{sec:conclusion}.

\section{Observation}
\label{sec:Observation}

Figure \ref{fig:box} shows 
the entire region of SNR \rxj\ observed with {\it Suzaku} as well as 
the {\it Chandra} image of the northwestern (NW) part of the remnant in the 0.5--10 keV band. 
The NW shell, located on the Galactic plane,  is the brightest part in the SNR. 

We have performed the {\it Chandra} ACIS-I observations of the NW shell of \rxj\ for 6 times; once in 2005, 2006 and 2011, and three times in 2009. There is also the initial pointing of the northwestern part in 2000 (see Table~\ref{tab:tab1}). All of the observations used in this paper are taken with the arrays of Advanced CCD Imaging Spectrometer (ACIS)-I on board the {\it Chandra} satellite, covering the field of view of 16.9\arcmin$\times$16.9\arcmin. The {\it Chandra} data are reprocessed by using {\tt chandra\_repro} with CALDB version 4.6.9 of CIAO version 4.7, the analysis software provided by the {\it Chandra X-Ray Center}\footnote{see http://cxc.harvard.edu}.
By combining the {\it Chandra} data over a span of 11 years, we measure the proper motion of the X-ray emitting shell. 

\begin{table}[h]
\tbl{{\it Chandra} data of  NW shell of \rxj}{%
\begin{tabular}{cccc}
\hline \hline
ObsID & Start Date &	Pointing position  &  Exposure \\
 	&	& [$\alpha_{\rm J2000}, \delta_{\rm J2000}$] & [ks]  \\ \hline 
12671  &    2011-07-01 &  \timeform{17h11m47.5s}, \timeform{-39D33m41.2s}   &      89.9   \\   
10092  &    2009-09-10   &  \timeform{17h11m46.1s}, \timeform{-39D33m51.6s}   &      29.2   \\   
10091  &    2009-05-16   &  \timeform{17h11m46.3s}, \timeform{-39D32m55.7s}   &      29.6   \\   
10090  &    2009-01-30   &  \timeform{17h11m44.4s}, \timeform{-39D32m57.1s}   &      28.4   \\   
6370  &    2006-05-03    &  \timeform{17h11m46.6s}, \timeform{-39D33m12.0s}   &      29.8   \\   
5560  &    2005-07-09   &  \timeform{17h11m45.5s}, \timeform{-39D33m40.0s}   &      29.0   \\   
736 $^{\rm a}$  &    2000-07-25   &  \timeform{17h11m49.9s}, \timeform{-39D36m14.7s}    &  29.6   \\   
\hline \hline
\end{tabular}}
\label{tab:tab1}
 \begin{tabnote}
     $^{\rm a}$ \ PI: P. Slane. (PI: Y. Uchiyama for the other observations.)
 \end{tabnote}
\end{table}

\section{Analysis and Results}
\label{sec:sec3}

\subsection{Flux image}
\label{sec:sec3.1}
We produce flux images taking the spectral shape into consideration. Flux images divided into three energy ranges, 0.5$-$1.2 keV, 1.2$-$2 keV and 2$-$5 keV, are generated using {\tt fluximage} of CIAO. These three flux images are then added together to produce 0.5$-$5 keV flux images. The bin size of the flux image is set to be $1.968\arcsec$. 

Since \rxj\ is an extended source covering a large portion of each {\it Chandra} FoV, 
 it is difficult to accurately estimate the level of the background directly from the observation data.
 Therefore we discard photons with energies greater than 5 keV 
so that the uncertainties of the background become negligible. 
We do not subtract the background (non-X-ray background and astrophysical background emissions) from the flux images.

\begin{figure}[ht]
 \begin{center}
 \includegraphics[width=8cm]{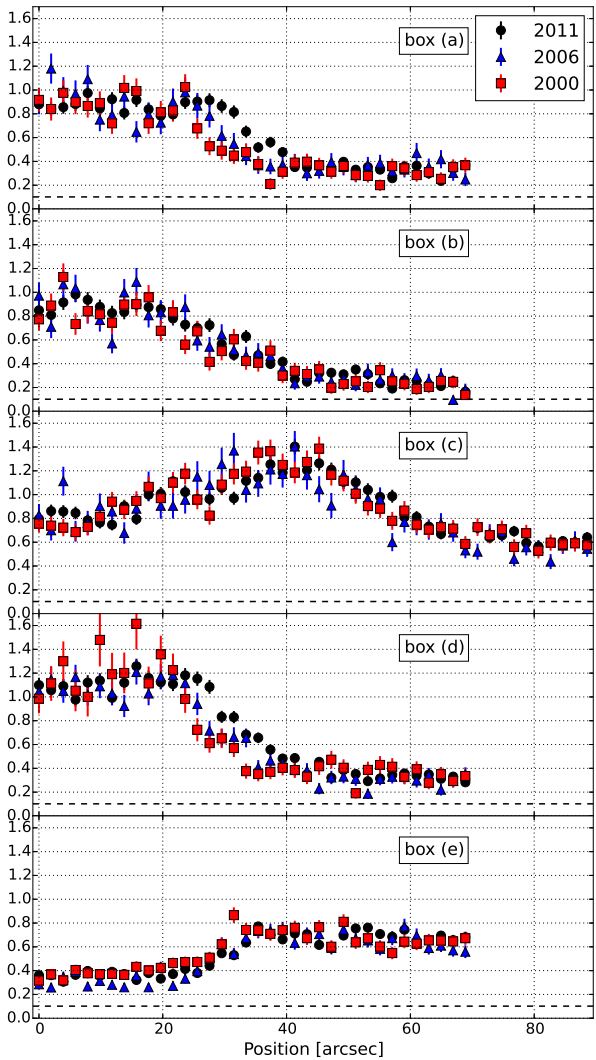}
 \end{center}
 \caption{Projection profile of box (a) to box (e) --- The black, blue and red plots refer to the data taken in 2011, 2006 and 2000, respectively. The horizontal dashed line shows the value of the faint region represented by box (f) taken in 2011. The vertical axis is in units of $10^{-7}\ {\rm photons}/{\rm cm}^2/{\rm s}/{\rm arcsec}$.}
 \label{fig:proj}
\end{figure}

\subsection{Expansion measurement}
\label{sec:ExpansionMeasurement}
Thanks to the superb angular resolution of {\it Chandra}, the X-ray emitting structures in \rxj, such as the outer or inner edges and the filaments in the shell, are clearly visible. We define the boxes along these spatial structures as shown in Fig.~\ref{fig:box} (right panel) in order to measure the proper motions of these relatively well-defined structures.  Proper motion measurements for boxes (a) and (e) are intended to measure the expansion of the remnant, as the edge structures in these two boxes are perpendicular to the radial direction that is the expected direction of the proper motion.  On the other hand, boxes (b), (c) and (d) are used  to measure the proper motions of the structures of interior regions although they are not perpendicular to the radial direction. The box size is as follows:  69\arcsec$\times$89\arcsec \ for box (a);  69\arcsec$\times$79\arcsec \ for box (b);  89\arcsec$\times$118\arcsec \ for box (c);  69\arcsec $\times$98\arcsec \ for box (d); and  89\arcsec$\times$188\arcsec \ for box (e). We checked that the choice of the orientation of the box does not affect our results as long as it is within $\pm 5^{\circ}$ from the chosen position angle. Box (f) is the representative of the faint regions outside the SNR, with the size of 1\arcmin$\times$1\arcmin. 

The flux images are not adjusted to compensate for possible registration error. As described in Section \ref{sec:errors}, 
we find that the relative registration errors of the images taken at different times are about 0.5 arcsec, which has only minor effects 
on the proper-motion measurements.

The one dimensional profiles of boxes (a)--(e) are shown in Fig.~\ref{fig:proj}, where the value at each flux-image pixel  
is summed over the direction along the longer side of the box.
The error for each spatial bin, $\sigma_i$, is calculated as
\begin{eqnarray}
\sigma_i = \frac{f_i}{\sqrt{N_i}} \; ,
\label{eq:sigma}
\end{eqnarray}
where $i$, $f_i$ and $N_i$ indicate the bin of the projection profile, the summed flux, and the summed photon counts, respectively. 
The dashed horizontal line shows a typical background level that includes particle as well as diffuse X-ray backgrounds,  
$1.0\times10^{-8}\ {\rm photons}/{\rm cm}^2/{\rm s}/{\rm arcsec}$, determined from box (f). 
In Fig.~\ref{fig:proj}, we can clearly see the displacements of the edge positions with time 
for boxes (a) and (d), while such  displacements,  if any, appear smaller for the other boxes.

In order to quantify the displacement of a projection profile with respect to the last-epoch
 (the year 2011) pointing that has the longest exposure time, 
 we search the best-matched shift position in the following way (see e.g., \cite{Katsuda2008,Katsuda2010}). 
By treating  the last-epoch profile as ``model",  we shift the model profile and 
compare it with a profile in another epoch. Then we calculate the $\chi^2$ values between the model and the profile in this epoch, 
where $\chi^2$ is given by 
\begin{eqnarray}
\chi^{2} =  \sum_{i} \frac{(f_i-m_i)^2}{\sigma_i^2}  \; , 
\label{eq:chi-sq}
\end{eqnarray}
with $m_i$  being the model profile value at a shift position. 

\begin{figure}[h]
 \begin{center}
 \includegraphics[width=8cm]{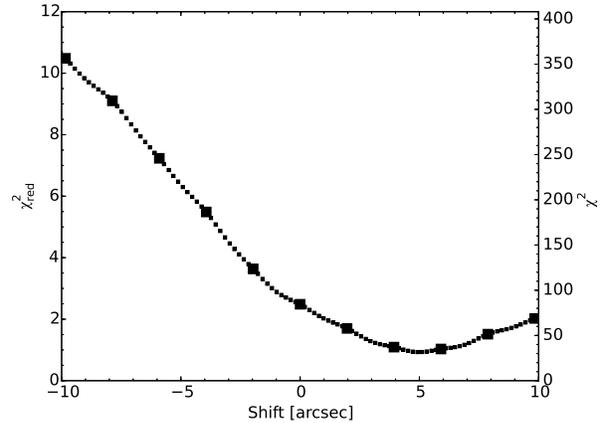}
 \end{center} 
\caption{$\chi^2$ distribution as a function of shift position obtained for box (a) in 2006. 
The large and small points show the results of using the actual data and the interpolated data, respectively. The positive shift indicates the radially outward direction which we expect is the expansion direction. The degree of freedom is 34 in this case.} 
\label{fig:chi2}
\end{figure}

Figure \ref{fig:chi2} shows the reduced $\chi^2$ ($\chi^2_{\rm red}$), i.e., $\chi^2$ divided by the degree of freedom, obtained for the profile of box (a) in the year 2006 as a function of shift position. 
Applying spline interpolation to the model profile, we determine the best-shift position 
 which gives the minimum value of $\chi^2_{\rm red}$. 
 In the case of box (a) in 2006, we find that the edge-like structure moved by 4.92\arcsec \ from 2006 to 2011.

\begin{figure}[ht]
 \begin{center}
 \includegraphics[width=8cm]{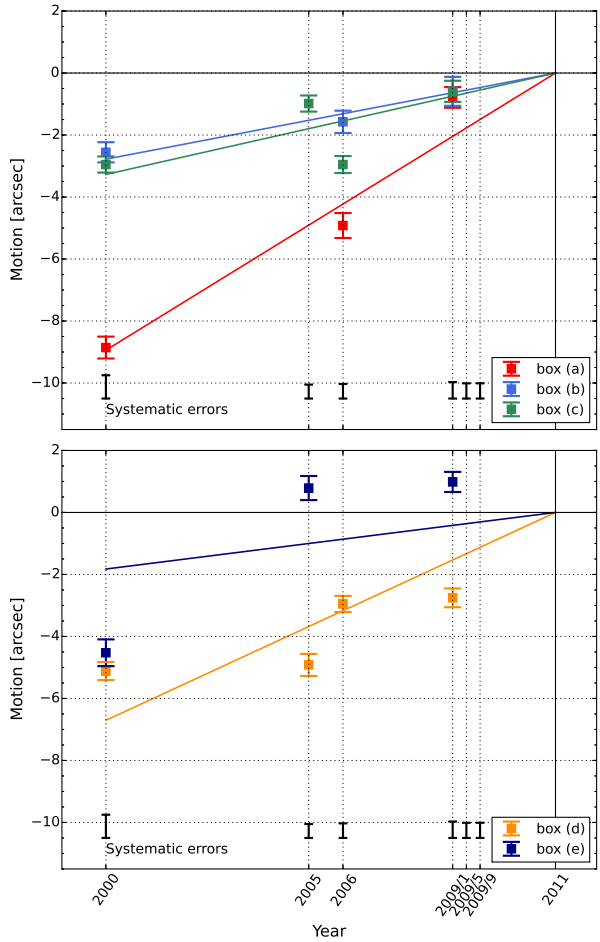}
 \end{center} 
\caption{Position shifts measured for boxes (a)--(e)  as a function of time, which reveal 
proper motions.  The systematic errors are shown at the bottom of each panel. We omit the data that exhibit significant variability} and the ones in May and September 2009.
\label{fig:motion}
\end{figure}

The measured proper motions obtained for boxes (a)--(e) are shown in Fig.~\ref{fig:motion}.
As mentioned above, the proper motions for boxes (a) and (d) are significantly larger than those for the other boxes. In the following two sections,  \S\ref{sec:errors} and  \S\ref{sec:fluxV}, we discuss uncertainties involved in the proper-motion measurements.

\subsection{Statistical and systematic errors}
\label{sec:errors}
Now we discuss statistical and systematic errors in Fig.~\ref{fig:motion}.
The statistical errors come from the $\chi^2$ fitting. The best-shift position is obtained when the $\chi^2$ value reaches 
the minimum, $\chi^2_{\rm min}$. 
The $1\sigma$ statistical error is set as the range of position shift that 
satisfies $\chi^2 \leq \chi^2_{\rm min} + 1.0 $.

The systematic errors are supposed to come from the absolute positional uncertainties of source coordinates in the {\it Chandra} observations. Nominal uncertainty of the registration is
known as 0.6\arcsec\ at 90\% confidence level\footnote{http://cxc.harvard.edu/cal/ASPECT/celmon/}. 
To confirm whether the registration of our {\it Chandra} data is accurate at this level, 
we checked  misalignments of field point-like sources, which can be regarded as being fixed on the sky for our purpose. 
 Then the misalignments of these point-like sources can be taken as the observation-specific systematic error. 
 We detect point-like sources by using {\tt wavdetect} of CIAO and picked up 10 useful 
 sources  based on the criteria 
that they are bright enough and located within 6\arcmin \ from the observation pointing center. Note the point spread function is worse 
at large off-axis angles, resulting in large uncertainty of the detected point-like source position. 

Setting the coordinates of the point-like sources detected in 2011 as fiducial values, the systematic error can be obtained by calculating RMS (root mean square) of misalignments between the fiducial coordinates and source coordinates detected in other observations. The average of these RMS values is 0.53\arcsec; the maximum one is 0.75\arcsec\ in 2000 and the minimum one is 0.45\arcsec\ in 2005. The average value of RMS is consistent with the 
known positional accuracy in {\it Chandra} observations.

\subsection{Flux variation}
\label{sec:fluxV}
Although we defer the detailed analysis of  flux variability \citep{Uchiyama2007} 
to a future publication, we have to take the effect of the flux variation into consideration when we perform the proper-motion measurements. 
The method employed in Section \ref{sec:ExpansionMeasurement} presumes 
a time-independent shape of projection profiles. This assumption is not valid anymore if 
 there is a significant flux variability.

\begin{figure}[ht]
 \begin{center}
 \includegraphics[width=8cm]{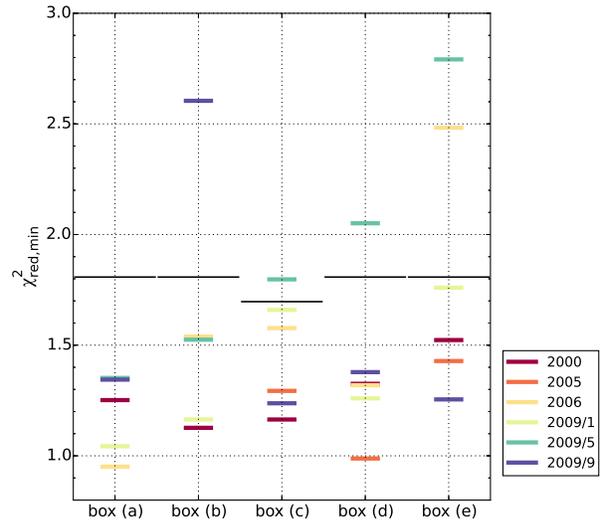}
 \end{center} 
\caption{Minimum reduced chi-squared values for each box and 
for each observation. The black horizontal lines are the 3$\sigma$ significance levels for flux variability.} 
\label{fig:min_redchi}
\end{figure}

In Fig.~\ref{fig:min_redchi} we plot the minimum $\chi^2_{\rm red}$ values for each box and for each epoch. 
The horizontal solid black line represents a $3\sigma$ significance level 
for the presence of time variability. For each box, we do not use epochs that exhibit significant variability when calculating the angular velocity in \S\ref{sec:angV}. 

\subsection{Angular velocity}
\label{sec:angV}

We can derive the angular velocity of the proper motion by fitting the shift positions in Fig.~\ref{fig:motion} with a straight line. 
Assuming  the distance to \rxj\ is 1 kpc \citep{Fukui2003}, 
we convert the angular velocity into the  velocity in units of km per second (Table~\ref{tab:tab2}).

\begin{table}[h]
\tbl{Proper motion measurement}{%
\begin{tabular}{p{2cm}p{2.5cm}p{2.5cm}}
\hline \hline
	   & Angular velocity 	& Velocity  \\ 
box ID	& [\arcsec/yr]  		& [km/s]   \\ \hline
box (a)  &  0.82 $\pm$ 0.06  &  3900 $\pm$ 300 \\     
box (b)  &  0.25 $\pm$ 0.06  &  1200 $\pm$ 300 \\     
box (c)  &  0.3 $\pm$ 0.05  &  1400 $\pm$ 200 \\     
box (d)  &  0.61 $\pm$ 0.05  &  2900 $\pm$ 200 \\     
box (e)  &  0.17 $\pm$ 0.06  &  800 $\pm$ 300 \\ 
\hline 
\end{tabular}}
\label{tab:tab2}
 \begin{tabnote}
The velocity uncertainties refer to one standard deviation errors.
The distance to \rxj\ is assumed to be 1 kpc \citep{Fukui2003}. 
 \end{tabnote}
\end{table}

Boxes (a) and (d) exhibit fast proper motions, 
3900 $\pm$ 300 km s$^{-1}$ and  2900 $\pm$ 200 km s$^{-1}$, respectively
(Table~\ref{tab:tab2}). 
Note that other boxes may have significant flux variability, making 
the result of the angular velocity in these cases less reliable.
It also should be noted that the variation of the obtained proper-motion velocities in Table~\ref{tab:tab2} can stem from the effect of the projection along the line of sight 
and/or interactions with the inhomogeneous medium (e.g.\ with dense clouds). 
Collisions with dense clouds can produce secondary shock waves with a reduced shock speed. 
The proper-motion velocities are significantly different between box (c) and (d) although these two boxes seem to be located on the same filamentary structure. This may be because box (d) is moving inside a void-like region with very low density gas \citep{Uchiyama2003}, resulting in a faster propagation speed.

Since box (a) appears to be located at the outer boundary of \rxj, the outer shock wave is supposed to be present at this position. 
Also, the proper motion measurement for box (a) is likely  most reliable 
given the reasonable $\chi^2_{\rm red, min}$. For these reasons, we assume the proper motion speed of box (a), 3900 $\pm$ 300 km s$^{-1}$,  corresponds to the blast-wave shock velocity of \rxj\ in the NW shell. In \S\ref{sec:Models} and \S\ref{sec:discussion}, we discuss the evolution models based on this result, assuming the velocity found 
in box (a) is the representative of the outer blast-wave shock speed. 


\section{Models}
\label{sec:Models}
The evolution of nonradiative SNRs, from ejecta-dominated (ED) phase to Sedov-Taylor 
(ST) phase, can be expressed by a single unified solution (\cite{TM99}). In this section, we investigate the parameters of SNR \rxj\ by combining the analytic solution of SNR evolution models with the observed quantities. 
We assume that the interactions with molecular clouds \citep{Fukui2012} 
have not yet affected the motion of 
the main blast wave of the remnant, and 
the evolution of \rxj\ can be described by the solution for 
spherically symmetric and nonradiative SNRs.

\subsection{Expanding in homogeneous ISM}
\label{sec:secTM}

\citet{TM99} described the evolution of spherically symmetric and nonradiative SNRs  from the 
ED stage to the ST stage. In particular, the case of the homogeneous ambient medium was investigated in detail. 
The density profiles of the ejecta and the ambient medium are written in the form of power law,
\begin{eqnarray}
\rho_{{\rm ej}} &\propto& r^{-n}   ,
\label{eq:modelEj} \\
\rho_{{\rm amb}} &\propto& r^{-s} , 
\label{eq:model}
\end{eqnarray}
respectively, where $r$ is the radius from the explosion center. In the case of $s=0$ which describes the ejecta expanding into the uniform interstellar medium, we define the ambient density profile as 
\begin{eqnarray}
\rho_{{\rm amb}}(s=0) &=& n_0 \; \mass ,
\end{eqnarray}
where $n_0$ and $\mu_{\rm H}$ are the hydrogen number density in units of ${\rm cm}^{-3}$ and the mean mass per hydrogen assuming cosmic abundance, respectively. 
We use 
$\mu_{\rm H} = 1.4\, m_{\rm H}$ with $m_{\rm H}$ being the mass of hydrogen.

For $n > 5$, the ejecta is devided into two portions consisting of {\it core} with uniform density and {\it envelope} with power-law density profile in the framework of \citet{TM99}. It should be noted that the physical behavior depends on whether the reverse shock is in the core or in the envelope. Thus the early stage of SNR evolution can be classified into three stages; ejecta-dominated stage with the reverse shock in the envelope (hereafter, ED-envelope stage); ED stage with the reverse shock in the core (ED-core stage); and Sedov-Taylor stage with the reverse shock in the core (ST stage). The model with $n$ larger than 5 is known to express the behavior of SNRs well. For instance, the model with $n=7$ would be appropriate for a typical Type Ia supernova \citep{TM99, Chevalier}, and the model with $n=9\mbox{--}12$ represents a typical type I\hspace{-.1em}I supernova \citep{Matzner1999}.

\citet{TM99} obtained the approximate analytic solutions of the motions of the  two shocks: the blast-wave shock and the reverse shock. To express these hydrodynamic properties of the motions of the shocks, we use a dimensionless form determined by characteristic values. The characteristic scales of length and time are
\begin{eqnarray}
R_{\rm ch}  &=& 3.07\left( \frac{M_{\rm ej}}{M_{\odot}} \right)^{1/3} n_0^{-1/3}  \; {\rm pc} \; , \\  
t_{\rm ch}  &=& 423 \left( \frac{E_{\rm ej}}{10^{51}\ {\rm erg}} \right)^{-1/2} \left( \frac{M_{\rm ej}}{M_{\odot}} \right)^{5/6} n_0^{-1/3}  \; {\rm yr}  \; ,
\end{eqnarray}
respectively. These scales are defined by using the initial parameters: the mass of ejecta ($M_{\rm ej}$), the kinetic energy of ejecta ($E_{\rm ej}$) and the density of the ambient medium ($n_0$). The dimensionless variable, for instance $F^*$, is expressed as $F^*=F/F_{\rm ch}$.

\begin{figure}[htb]
 \begin{center}
 \includegraphics[width=8cm]{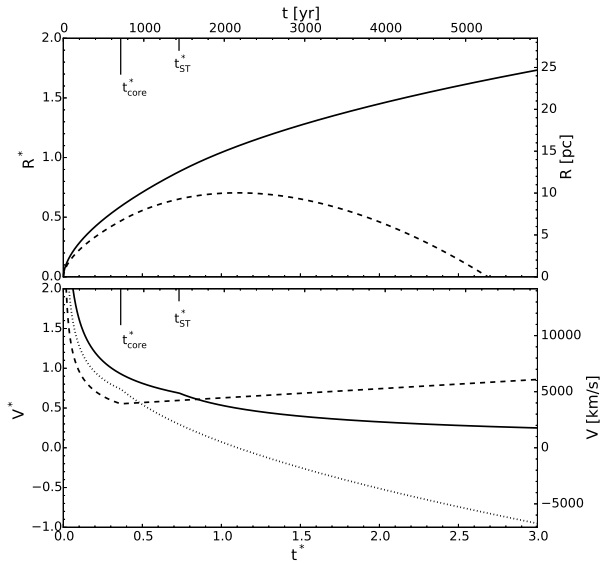} 
 \end{center}
\caption{Trajectories of the parameter set with ($s,\ n,\ M_{\rm ej},\ E_{\rm ej},\ n_0$) $=( 0,\ 7,\ 1\ M_{\Sol},\ 10^{51}\ {\rm erg},\ 0.01\ {\rm cm}^{-3})$. {\it Top panel}: Shock positions for the blast-wave shock (solid line) and the reverse shock (dashed line). {\it Bottom panel}: Shock velocities for the blast-wave shock (solid line), the reverse shock velocity at the pre-shock ejecta frame (dashed line), and the reverse shock velocity at the ambient rest frame (dotted line).} 
\label{fig:traj0}
\end{figure}

Once the parameter sets to determine the density profile and the initial conditions, i.e.\ ($s, n, M_{\rm ej}, E_{\rm ej}, n_0$), are chosen, one can obtain the trajectories of the shock motions, which are the profiles of the position of the blast-wave shock ($R_b$), its velocity ($v_b$), the position of the reverse shock ($R_r$) and its velocity ($v_r$) as a function of time. The shock velocities $v_b$ and $v_r$ are defined as the velocities at the ambient rest frame. One example of trajectories with the parameter set of ($s,\ n,\ M_{\rm ej},\ E_{\rm ej},\ n_0$) $=( 0,\ 7,\ 1\ M_{\Sol},\ 10^{51}\ {\rm erg},\ 0.01\ {\rm cm}^{-3})$ is shown in Fig.~\ref{fig:traj0}.

\subsection{Expanding in wind-blown bubble}
Following the work done by \citet{TM99} for deriving the analytic solution of SNR evolution especially in the case of $s=0$, \citet{LH03} extended the analytic solutions to the case of $s=2$. Note that there is a few minor revisions to Appendix of \citet{LH03}, reported in \citet{Hwang2012}. 

The analytic solutions of SNR evolution developed by \citet{LH03} 
are relevant if a SNR is expanding into a cavity created by a circumstellar wind from the progenitor. The ambient medium density profile can be defined as,
\begin{eqnarray}
\rho_{{\rm amb}}(s=2) = n_2 \left( \frac{r}{R_b} \right)^{-2} \;  \mass ,
\end{eqnarray}
where $n_2$ indicates the hydrogen number density at the position of the blast-wave shock in units of ${\rm cm}^{-3}$.
In this model, we assume that the profile can be described as $\propto r^{-2}$ at any distance from the explosion center and the shock has not yet reached the cavity wall. 

\begin{figure}[htb]
 \begin{center}
 \includegraphics[width=8cm]{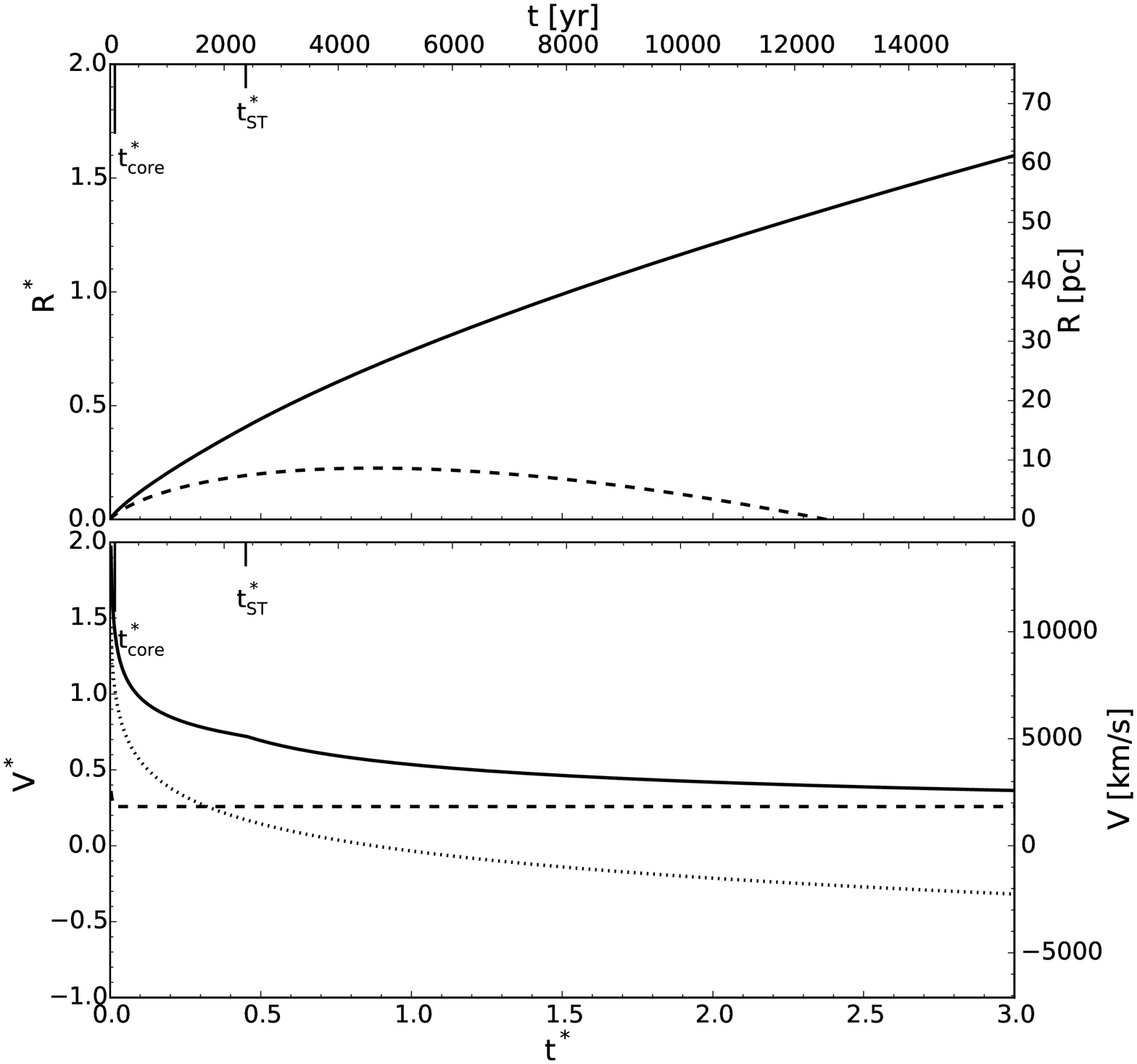} 
 \end{center}
\caption{Trajectories of the parameter set with ($s,\ n,\ M_{\rm ej},\ E_{\rm ej},\ n_2$) $=( 2,\ 7,\ 1\ M_{\Sol},\ 10^{51}\ {\rm erg},\ 0.01\ {\rm cm}^{-3})$. Each line denotes the same quantity as Fig.~\ref{fig:traj0}.} 
\label{fig:traj2}
\end{figure}

The characteristic scales for the $s=2$ case are expressed as
\begin{eqnarray}
R_{\rm ch} &=& 28.9 \left( \frac{M_{\rm ej}}{M_{\odot}} \right) \left( n_2 R_{b, {\rm pc}}^2 \right)^{-1} \; {\rm pc} , \\
t_{\rm ch} &=& 3987 \left( \frac{E_{\rm ej}}{10^{51} {\rm erg}} \right)^{-1/2} \left( \frac{M_{\rm ej}}{M_{\odot}} \right)^{3/2} \left( n_2 R_{b, {\rm pc}}^2 \right)^{-1}  \; {\rm yr} ,
\end{eqnarray}
where $R_{b, {\rm pc}}$ is the blast-wave shock radius in units of pc.
One example of trajectories of the $s=2$ case is shown in Fig.~\ref{fig:traj2}, with the parameter set of ($s,\ n,\ M_{\rm ej},\ E_{\rm ej},\ n_2$) $=( 2,\ 7,\ 1\ M_{\Sol},\ 10^{51}\ {\rm erg},\ 0.01\ {\rm cm}^{-3})$.

It is reasonable to assume that \rxj\ exploded inside a wind-blown bubble (e.g., \cite{Berezhko2010}). There is a neutron star candidate around the center of \rxj\ 
in projection, implying a core-collapse supernova. 
In addition, the relatively thick shell of the western part of \rxj\ \citep{Hiraga2005} favors 
a decreasing ambient density profile such as $\rho \propto r^{-2}$ which 
can be created by a stellar wind. 
We adopt the circumstellar wind model with $s=2$, but present 
the uniform interstellar medium model with $s=0$ for comparison.

\subsection{Constraints from the observational data} 
\label{sec:analysis}

We explore the evolution models by varying model parameters in order to 
find the parameters that can describe the observed properties of \rxj. 
We use the measurement of the blast-wave radius and velocity (\S\ref{sec:angV}) 
as well as observational constraints on the ambient density and the ejecta mass.
The distance to \rxj, $d$, is well constrained by the association with the 
adjacent molecular clouds \citep{Fukui2003, Fukui2012,Sano2013}; 
we adopt $d=1$ kpc \citep{Fukui2003}  in this paper. 

First, we fix the current value of the blast-wave shock radius to be $R_b=8.68$ pc, 
corresponding to the location of box (a). 
 The blast-wave shock speed in the NW region is measured as 
 $(3900 \pm 300) (d/{\rm kpc})  \; {\rm km}\ {\rm s}^{-1}$ (\S\ref{sec:angV}).

We set the initial parameters, $M_{\rm ej}$, $E_{\rm ej}$ and $n_s$, in
plausible ranges. We take $E_{51} = 0.5$, 1.0 and 2.0 as the ejecta kinetic energy in units of $10^{51}$ erg. 
For the ejecta mass, we set a range of $M_{\rm ej} = 0.6- 10 \ M_{\Sol}$, 
given that the X-ray emitting ejecta mass is estimated as $0.63 - 0.8 \ M_{\Sol}$ and 
the progenitor mass is likely to be less than 20 $M_{\Sol}$ from the recent X-ray line 
measurements \citep{Katsuda2015}.

Previous X-ray and gamma-ray observations demonstrated that the ambient density should be low. From the X-ray observations, the lack of the thermal X-ray emission in the western shell gave the upper limit to the number density of the ambient medium: $0.02\, (d/{\rm kpc})^{-1/2} \ {\rm cm}^{-3}$ with the {\it XMM-Newton} satellite 
\citep{Cassam2004} and $0.01\, (d/{\rm kpc})^{-1/2} \ {\rm cm}^{-3}$ with  the {\it Suzaku} satellite \citep{Takahashi2008}. 
Here we assume that the electron temperature is about 1 keV, 
as expected for the shock speed of $\sim 4000 \; {\rm km}\ {\rm s}^{-1}$ 
(\S\ref{sec:angV}). 
Based on the constraints on the ambient density from the X-ray observations, we set the range of $n_s$ to be $0.002 - 0.02\ {\rm cm}^{-3}$.

It should be noted that the blast-wave shock velocity, which we deduced from 
the proper motion measurement  in \S\ref{sec:ExpansionMeasurement}, 
also indicates the low ambient density. 
Generally, the blast-wave velocity in the ST stage is given by
\begin{eqnarray}
v_b = \frac{2}{5-s} \left( \xi_s \frac{E_{\rm ej}}{\rho_{\rm amb}} \right)^{1/2} R_b^{(s-3)/2} \; ,
\label{eq:Vb}
\end{eqnarray}
where $\xi_s = \sqrt{(5-s)(10-3s)/8\pi}$. 
Using  $R_b = 8.68$ pc and $v_b = 3900\pm300 \; {\rm km}\ {\rm s}^{-1}$,  
Equation.~(\ref{eq:Vb}) leads to $0.027 E_{51} \leq n_2 \leq 0.036 E_{51}$ for $s=2$, and $0.048 E_{51}  \leq n_0 \leq 0.056 E_{51}$ for $s=0$. If the SNR is in the ED stage, the ambient density must be lower than these densities derived from the ST solution. Our measurement of the blast-wave shock speed sets the upper limit to the ambient medium density as  0.036 ${\rm cm}^{-3}$ for $E_{\rm ej}=10^{51}$ erg in the case of $s=2$.

Using the analytic solutions given in \citet{TM99} and \citet{LH03}, we can obtain the physical values of ($ t_{\rm age},\ v_b,\ R_r,\ v_r, \ M_b, \ M_r,$), assuming the parameter sets of ($s,\ n,\ M_{\rm ej},\ E_{\rm ej},\ n_s$) and 
 $R_b = 8.68\ {\rm pc}$. Here $t_{\rm age}$ is the current age of the SNR. $M_b$ and $M_r$ are the blast-wave-shocked ambient mass and the reverse-shocked ejecta mass (see \S\ref{sec:shocked_ejecta} for details), respectively.

One set of the parameters leads to one set of the dimensionless trajectories (see Fig.~\ref{fig:traj0} and Fig.~\ref{fig:traj2}). First, the time which gives $R_b=$8.68 pc, i.e.\ the present age of the SNR ($t_{\rm age}$), can be determined directly from the trajectory of the blast-wave position. Then the values of $v_b$, $R_r$, $v_r$, $M_b$ and $M_r$ at $t=t_{\rm age}$ are uniquely calculated.  
Comparing $t_{\rm age}$ with $t_{\rm ST}$ and $t_{\rm core}$, we can obtain the evolutional phase which the SNR is in, where $t_{\rm ST}$ and $t_{\rm core}$ denote the transition time from the ED-core stage to the ST stage and the transition time from the ED-envelope stage to the ED-core stage, respectively.

\begin{figure}[ht]
\begin{center}
 \includegraphics[width=8.5cm]{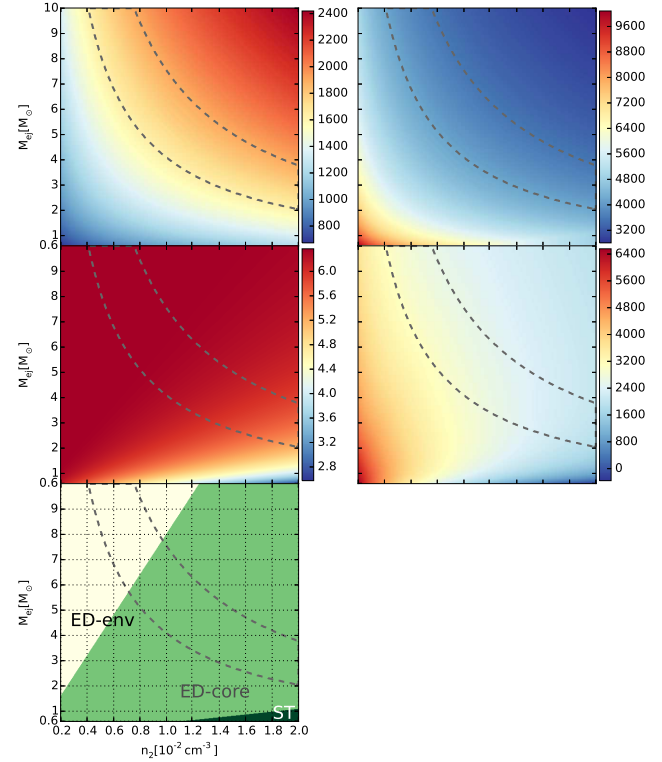}
 \end{center}
\caption{The present-day SNR age and blast wave velocity 
as functions $n_2$ and $M_{\rm ej}$ for the parameter set $(n, s, E_{51}) = (7, 2, 1)$. The x axis and y axis represent $n_2$ in units of $10^{-2} {\rm cm}^{-3}$ and $M_{\rm ej}$ in units of $M_{\Sol}$, respectively. The upper left panel shows two dimensional image of $t_{\rm age}$ in units of yr which satisfy $R_b = 8.68\ {\rm pc}$. The upper right one, middle left one, middle right and lower left one illustrate that of $v_b$ in units of km/s, $R_r$ in units of pc, $v_r$ in units of km/s, and the evolutional phase at $t=t_{\rm age}$, respectively.} 
\label{fig:LH}
\end{figure}

We present the case with $(s, n, E_{\rm ej, 51}) = (2, 7, 1)$ as a characteristic example. 
Figure~\ref{fig:LH} illustrates two dimensional images of $t_{\rm age}$ (upper left), $v_b$ (upper right), $R_r$ (middle left), $v_r$ (middle right) and the evolutional phase (lower left) for  the ambient density of $0.002 - 0.02\ {\rm cm}^{-3}$ ($x$ axis) and the ejecta mass of $0.6 - 10\ M_{\Sol}$ ($y$ axis). All images are overlaid with the range of the blast-wave shock velocity ($v_b = 3900 \pm 300 \ {\rm km/s}$) using dashed lines. 
Some points inside the dashed-lined region are selected and their parameter sets are listed in Table~\ref{tab:tab3}, discussed in \S\ref{sec:discussion}.

\section{Discussion}
\label{sec:discussion}

We explore 210,000 parameter sets for ISM ($s=0$) and CSM ($s=2$) cases. The range of the ejecta mass $M_{\rm ej} = 0.6\mbox{--}10 M_{\Sol}$ is divided into 100 grids, and the range of the ambient density $n_s = 0.002 \mbox{--} 0.02\ {\rm cm^{-3}}$ into 100 grids. The ejecta energy is $E_{51} =$ 0.5, 1, 2, and the ejecta density profile index is $n = 6$, 7, 8, 9, 10, 12, and 14. 
Similar analysis on the evolution model for SNR RX~J0852.0$-$4622 (Vela Jr.) 
was performed by \citet{Allen2015}. 
In Table~\ref{tab:tab3}
we show the values of $M_{\rm ej}$ and $n_s$ that can 
reproduce the blast-wave velocity in the cases of $n=7,\ 9$ and $E_{51}=1$.

\begin{table*}[ht]
\tbl{Evolution model for $E_{51}=1$.}{
\begin{tabular}{cccccccccccccc}
\hline \hline
Model  &  $s$ & $n$  &  age &  $v_b$  &  $R_r$ &  $v_{r, {\rm obs}}$ &  $n_s$   &  $M_{\rm ej}$	  &  $E_{\rm ej}$  & $t_{\rm ST}$ & $t_{M_b = M_{\rm ej}}$   & $M_b$  &    $M_r$    \\ 
 &   &   &  [yr] &  [km/s]  &  [pc] &  [km/s] &  [${\rm cm}^{-3}$]   &  [$M_{\Sol}$]  &  [$10^{51}$ erg]  & [yr] & [yr]   &  [$M_{\Sol}$]  &     [$M_{\Sol}$]    \\ \hline
 1  & 2  & 7 & 1620   & 4191 & 6.37 & 3074  &  0.005  & 8  & 1 & 107655 & 14015   &  1.42  & 1.72    \\
 2  & 2  &  7 & 1713   & 3964 & 6.32 & 2788  &  0.01  & 5  & 1 & 26596 & 3462   &  2.85  & 3.04    \\ 
 3  & 2  &  7 & 1668   & 4070 & 5.88 & 2389  &  0.015  & 3  & 1 & 8240 & 1072   &  4.27  & 2.52    \\ 
 4  & 2  &  7 & 1620   & 4191 & 5.14 & 1804  &  0.02  & 2  & 1 & 3364 & 437   &  5.7  & 1.87    \\ 
  \hline
 5  & 0  &  7 & 1199   & 4044 & 6.89 & 3209  &  0.008  & 8  & 1 & 8760 & 4737   &  0.76  & 0.38    \\ 
 6  & 0  &  7 & 1226   & 3954 & 6.89 & 3138  &  0.01  & 7  & 1 & 7276 & 3934   &  0.95  & 0.47    \\ 
 7  & 0  &  7 & 1180   & 4110 & 6.89 & 3262  &  0.015  & 4  & 1 & 3987 & 2156   &  1.42  & 0.71    \\ 
 8  & 0  &  7 & 1180   & 4110 & 6.89 & 3262  &  0.02  & 3  & 1 & 2850 & 1541   &  1.9  & 0.94    \\ 
  \hline \hline
 9  & 2  &  9 & 1932   & 3765 & 6.69 & 2899  &  0.005  & 10  & 1 & 53072  & 18781  &  1.42  & 3.1    \\ 
 10 & 2  &  9 & 2013   & 3613 & 6.61 & 2677  &  0.01  & 8  & 1 & 18987  &  6719   &  2.85  & 4.62    \\ 
 11  & 2  &  9 & 1842   & 3950 & 6.35 & 2696  &  0.015  & 5  & 1 & 6254  &  2213   &  4.27  & 3.79    \\ 
 12  & 2  &  9 & 1866   & 3900 & 6.16 & 2516  &  0.02  & 4.5  & 1 & 4005  &  1417   &  5.7  & 3.68    \\ 
  \hline
13 & 0  &  9 & 1522   & 3717 & 7.3 & 3124  &  0.005  & 10  & 1 & 8817  &  6988   &  0.47  & 0.44    \\ 
14  & 0  &  9 & 1517   & 3730 & 7.3 & 3134  &  0.01  & 7  & 1 & 5198  & 4120   &  0.95  & 0.88    \\ 
15  & 0  &  9 & 1451   & 3900 & 7.3 & 3277  &  0.015  & 5  & 1 & 3431  &  2719   &  1.42  & 1.32    \\ 
16  & 0  &  9 & 1413   & 4004 & 7.28 & 3256  &  0.02  & 4  & 1 & 2588  & 2051   &  1.9  & 1.69    \\ 
    \hline \hline
\end{tabular} }
\label{tab:tab3}
 \begin{tabnote}
  $R_b$ is fixed to 8.68 pc. $t_{\rm ST}$ and $t_{M_b = M_{\rm ej}}$ represent the transition time when ED solution connects Sedov solution smoothly and the nominal time when the blast-wave-shocked ambient mass becomes equal to the ejecta mass, respectively.
 \end{tabnote}
\end{table*}

\subsection{Age}
\label{sec:age}

The age of \rxj\ has been matter of debate. 
While \citet{Wang1997} proposed the connection between SN393 and SNR \rxj, \citet{Fesen2012} suggested that the historical descriptions about SN393 are not easily reconciled with the expected brightness and duration. 

\begin{figure}[h]
\begin{center}
 \includegraphics[width=7cm]{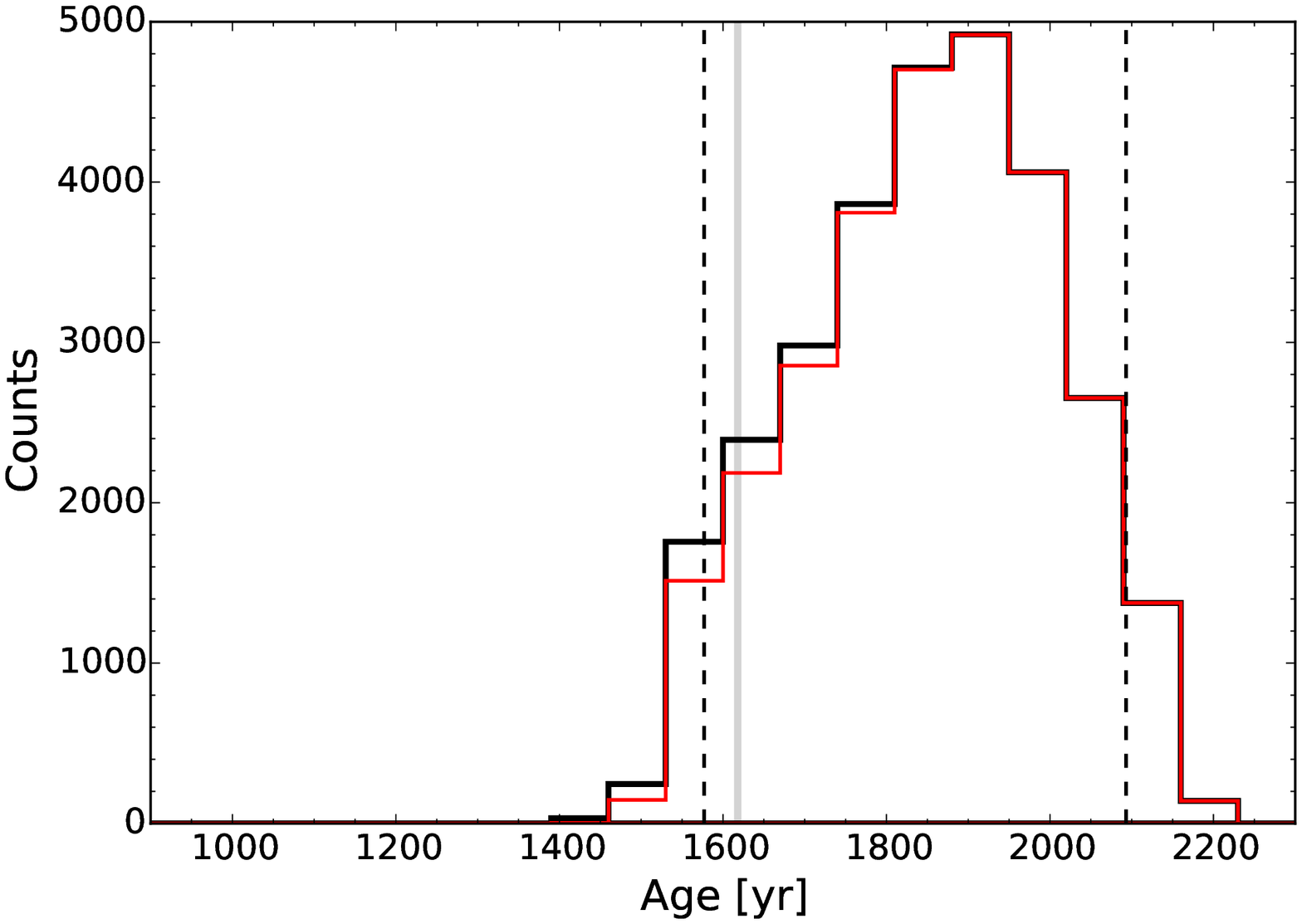} \\
 \includegraphics[width=7cm]{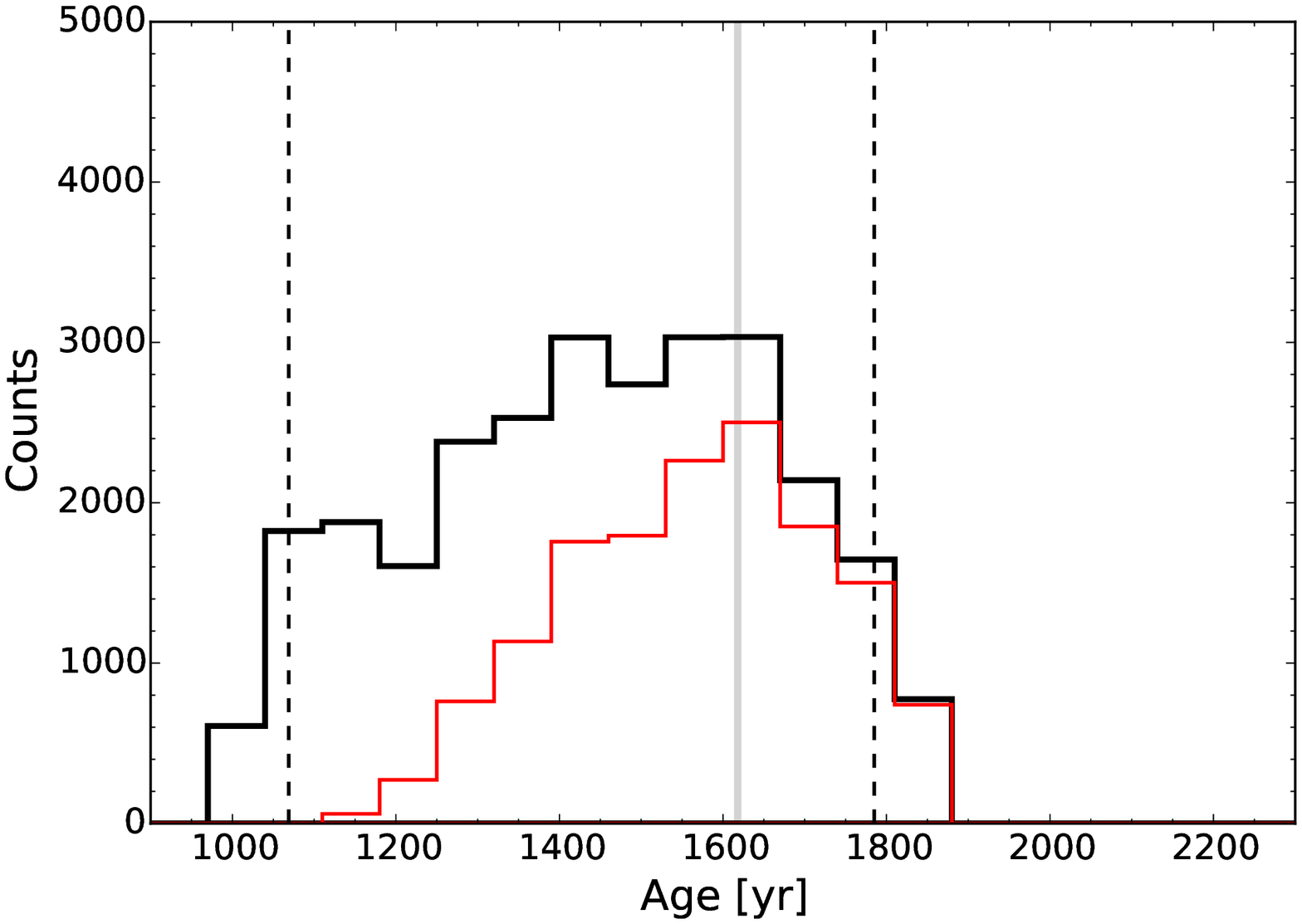}
   \end{center}
\caption{{\it Top panel}: Histogram (black) of the age ($t_{\rm age}$) for the $s=2$ case 
from the models that satisfy the blast-wave shock velocity of $3900\pm 300\, {\rm km}\, {\rm s}^{-1}$. The grey vertical line shows the age assuming the association with SN393. The dashed vertical lines present the 90\% confidence level range. 
The red histogram is constructed from the models that also satisfy a constraint of the 
mass of the shocked ejecta. 
{\it Bottom panel}: Same histogram for the $s=0$ case.} 
\label{fig:hist}
\end{figure}

Figure~\ref{fig:hist} presents the distributions of the SNR age, constructed 
from the parameter sets that agree with  the observed blast-wave shock velocity. 
We find $t_{\rm age} = 1580\mbox{--}2100$ years for $s=2$ and 
$t_{\rm age} = 1070\mbox{--}1790$ years for $s=0$ as a 90\% confidence interval 
assuming that each parameter set is equally probable (see also \cite{Allen2015}). 
These ranges are consistent with the age expected from the association with SN393:  $t_{\rm age} = 1618$ years for the fiducial year of 2011. 
Our age estimate depends only weakly on the distance, $d$, which is assumed to be 1 kpc in this paper. 
The uncertainty of the distance, up to $\sim 30\%$,  can change the age estimate by less than 5\%.

\begin{table}[h]
\tbl{Constraints on age and evolutional phase}{%
\begin{tabular}{ccc}
\hline \hline
	  & Age &  Phase   \\ 
  	  & [yr]  &  ED-env / ED-core / ST [\%]  \\ \hline 	
$s=2$ & 1580$-$2100 &  15.6 / 75.5 / 8.8 $^{\rm a}$ 	 \\ 
	   & 			 &  15.6 / 58.7 / 25.7 $^{\rm b}$  	\\ \hline
$s=0$& 1070$-$1790 &  71.8 / 28.2 / 0 $^{\rm a}$ 	  \\
	 & 		        &  71.8 / 25.2 / 3 $^{\rm b}$ 	  \\
\hline \hline
\end{tabular}}
\label{tab:tab5}
 \begin{tabnote}
    Age indicates the range at 90\% confidence level. Phase shows the model count ratio of ED-envelope stage, ED-core stage and ST stage.
    
    $^{\rm a}$ \ Adopting $t_{\rm ST}$ to the transition time.
    
    $^{\rm b}$ \ Adopting $t_{M_b = M_{\rm ej}}$ to the transition time.
 \end{tabnote}
\end{table}

\subsection{Shocked ejecta mass}
\label{sec:shocked_ejecta}

\citet{Katsuda2015} estimated the total X-ray emitting ejecta mass to be roughly $0.63\mbox{--} 0.8\, M_{\Sol}$ using the thermal X-ray line emissions. 
It is interesting to compare the mass estimate with the reverse-shocked ejecta mass ($M_r$) in our models, 
 which can be calculated by integrating the density profile of ejecta from the contact discontinuity position to the reverse shock position. 
However,  the X-ray emitting mass would not be taken as the total reverse-shocked ejecta mass, as also concerned in \citet{Katsuda2015}. Since the ejecta continues to expand after being shocked by the reverse shock, it is cooled 
by adiabatic expansion. 
A part of the ejecta being reverse-shocked in the past may have already been cooled and may not emit X-ray anymore at present. 
Still we can use $\sim 0.7\, M_{\Sol}$ as a lower limit to the total shocked ejecta mass. 
As can be seen in Table~\ref{tab:tab3}, the case of $s=0$ generally results in 
small $M_r$. 
In Fig.~\ref{fig:hist}, the distributions of the SNR age after imposing the constraint on 
the shocked ejecta mass are shown as  red histograms. 

\subsection{Evolutional phase}
\label{sec:evolutional_phase}
Now we discuss the evolutional phase to which SNR \rxj\ is currently belonging. 
There is some ambiguity of the definition of the transition time from the ED stage to the ST stage because the transition occurs gradually. 
 We adopt $t_{\rm ST}$ defined in \citet{TM99} for the $s=0$ case 
 and $t_{\rm conn}$ in \citet{LH03} as $t_{\rm ST}$ for the $s=2$ case. 
 We also consider a transition time when the blast-wave-shocked ambient medium mass ($M_b$) becomes equal to the ejecta mass. We express it as $t_{M_b = M_{\rm ej}}$. Note that $t_{M_b = M_{\rm ej}}$ is generally smaller than $t_{\rm ST}$. 

For most  of the parameter sets that are within the adopted range, 
the evolutional phase is the ED stage, especially if we use $t_{\rm ST}$ 
as the transition time (see Table~\ref{tab:tab5}). 
Contrary to what is widely considered, 
SNR \rxj\ is not in the ST stage. 
Because the total energy content of cosmic rays has to be large enough to  produce the observed gamma rays regardless of the emission models 
(see \cite{Edmon2011}), 
 a very young evolutional stage in Models 1, 2, 9, and 10 
where $t_{\rm age} \ll t_{\rm ST}$ would be inappropriate.

\subsection{Models for association with SN393}

Model 3 in Table~\ref{tab:tab3} represents one of the best parameter sets, 
being in good agreement with the observed blast-wave shock velocity, the estimated shocked ejecta mass and the realistic evolutional phase. 
Furthermore, the age is consistent with SN393. 
Note that the models with $s=0$ are hardly reconcilable with the observational 
constraints.  
The ambient density of $n_2 \sim 0.015\ {\rm cm}^{-3}$ is most plausible, being  consistent with the previous estimates. 
We find that the evolution of \rxj\ has not yet reached the ST stage. 
Its phase is still the ED stage or in the middle of the transition to 
the ST stage. 
 If indeed so, the highest energy particles at the blast wave do not reach 
 the maximum attainable energy yet. 

In Table~\ref{tab:tab4}, we show the evolution models for $s=2$ 
under the assumption that SN393 has produced \rxj, as the most likely scenario. 
An energetic SN with $E_{51}=2$ is difficult to be reconciled with the observational 
constraints, particularly the evolution phase; it is too young (i.e., $t_{\rm age} \ll t_{\rm ST}$) 
to accelerate enough amount of the cosmic-ray particles. 
On the other hand, a less energetic event with $E_{51}=0.5$ is also less likely 
than with $E_{51}=1$ because the revere-shocked mass is not large enough to 
produce the thermal X-ray emission \citep{Katsuda2015}.
Model 4 in  Table~\ref{tab:tab4}, which is quite similar to 
Model 3 in Table~\ref{tab:tab3},  is fully consistent with the various observational data.

\begin{table*}[ht]
\tbl{Evolution model under the assumption that SN393 produced \rxj. }{
\begin{tabular}{ccccccccccccccc}
\hline \hline
Model  &  $s$ & $n$  &  age & $R_b$ &  $v_b$  &  $R_r$ &  $v_{r, {\rm obs}}$ &  $n_s$   &  $M_{\rm ej}$	  &  $E_{\rm ej}$  & $t_{\rm ST}$ & $t_{M_b = M_{\rm ej}}$   & $M_b$  &    $M_r$    \\ 
 &   &   &  [yr] & [pc]  &  [km/s]  &  [pc] &  [km/s] &  [${\rm cm}^{-3}$]   &  [$M_{\Sol}$]  &  [$10^{51}$ erg]  & [yr] & [yr]   &  [$M_{\Sol}$]  &     [$M_{\Sol}$]    \\ \hline
1 & 2  &  7  &   1618  &  8.67  &  4193 &  5.13   &  1806  &   0.01   &  1.0   &   0.5  &   3364  &  437 &   2.85  & 0.94 \\
2 & 2  &  7  &   1618  &  8.75  &  3947 &  3.38   &  374  &   0.015   &  0.6   &   0.5  &   1042  &  135 &   4.27  & 0.59 \\
 \hline
3 & 2  &  7  &   1618  &  8.67  &  4193 &  6.25   &  2863  &   0.01   &  4.0   &   1.0  &   19030  &  2477 &   2.85  & 2.71 \\
4  & 2  &  7  &   1618  &  8.65  &  4182 &  5.78   &  2377  &   0.015   &  2.7   &   1.0  &   7035  &  916 &   4.27  & 2.32  \\
5  & 2  &  7  &   1618  &  8.67  &  4193 &  5.13   &  1806  &   0.02   &  2.0   &   1.0  &   3364  &  437 &   5.7  & 1.87 \\ 
  \hline
 7  & 2  &  7  &   1618  &  8.57  &  4142 &  6.27   &  2968  &   0.017   &  10.0   &   2.0  &   31290  &  4073 &   4.84  & 5.48 \\
 8  & 2  &  7  &   1618  &  8.67  &  4193 &  6.25   &  2863  &   0.02   &  8.0   &   2.0  &   19030  &  2477 &   5.7  & 5.43
 \\
    \hline \hline
\end{tabular} }
\label{tab:tab4}
 \begin{tabnote}
 \end{tabnote}
\end{table*}

\section{Conclusions}
\label{sec:conclusion}

We have measured the proper motions of the edge-like and filamentary structures visible in the NW shell of SNR \rxj. 
From the measurement for box (a) we deduced the shock velocity in NW region of the SNR as 
$(3900\pm 300) (d/{\rm kpc})\, {\rm km}\, {\rm s}^{-1}$, where the distance is 
estimated as $d \simeq 1$ kpc from CO observations \citep{Fukui2003}.
This relatively fast shock velocity is consistent with the presence of synchrotron X-ray emission 
(see \cite{Uchiyama2003}) and 
indicates that the ambient density is low, 
$n \sim 0.01\ {\rm cm}^{-3}$, being compatible with what has been suggested in the literature. 

Assuming that the shock speed we have measured with {\it Chandra} at the outermost edge of the NW shell is the representative of the remnant's outer shock wave as a whole, 
we have confronted our measurements as well as some previous results with 
 the analytic solution of the hydrodynamical properties of SNRs. 
We estimated the age of the remnant to be 1580 to 2100 years (a 90\% confidence interval) for the circumstellar wind model; 
the age estimate is almost unaffected by the relatively small uncertainty of the distance. 
Our X-ray measurements support the association with SN393 proposed by \citet{Wang1997} . 
A model with SN kinetic energy of $E = 10^{51}\ \rm erg$, the ejecta mass of 
$M_{\rm ej} = 2.7\, M_{\Sol}$, and the ambient density 
at the current blast wave location of $n_2 = 0.015\ {\rm cm}^{-3}$, provides reasonable 
explanation for the observed quantities. 
We stress that the transition to the ST phase is incomplete, $t_{\rm age} < t_{\rm ST}$, 
for any reasonable set of parameters. 
Generally, the maximum energy of accelerated protons is thought to reach its maximum at the beginning 
of the ST phase. Our results suggest that the current maximum energy would not correspond to the maximum attainable energy for this remnant.


\begin{ack}
We thank the anonymous referee for valuable comments on the manuscripts. 
We thank Taro Fukuyama for providing us with the {\it Suzaku} image. 
This work was supported in part by JSPS (Japan Society for the Promotion of Science) KAKENHI Grant Number JP26247027. 
Support for this work was provided by the National Aeronautics and Space Administration through Chandra Award Numbers GO9-0074X and  GO1-12092X issued by the Chandra X-ray Observatory Center, which is operated by the Smithsonian Astrophysical Observatory for and on behalf of the National Aeronautics Space Administration under contract NAS8-03060.
\end{ack}




\end{document}